\documentclass[twocolumn,prb,amssymb,showpacs,showkeys]{revtex4}
\usepackage{graphicx}
\usepackage{dcolumn}
\usepackage{bm}
\begin{document}
\title{Near Fermi level Electronic Structure of Pr$_{1-x}$Sr$_x$MnO$_3$:  
Photoemission Study}
\author{P. Pal, M. K. Dalai, R. Kundu, M. Chakraborty and B. R. Sekhar}
\email{sekhar@iopb.res.in}
\affiliation{Institute of Physics, Sachivalaya Marg, Bhubaneswar 751 005,
India.}
\author{C. Martin}
\affiliation{Laboratoire CRISMAT, UMR 6508, ISMRA, Boulevard du Marechal
Juin, 14050 Caen, France}
\begin{abstract}

In this study, we report the observation of a pseudogap associated with the 
insulator-metal transition in compositions of Pr$_{1-x}$Sr$_x$MnO$_3$ 
system with no charge ordering. Our valence band photoemission study shows 
that the observed shifts in the near Fermi level density of states are 
abrupt at the Curie transtion and occur over an energy scale of $\sim$ $1$ 
eV, strongly suggesting that the charge-ordering gap observed earlier in 
other manganites and the pseudogap observed here may indeed have same 
origin. These results could be understood within the framework of models 
based on electronic phase separation where it has been shown that the 
pseudogap is a generic feature of the mixed-phase compositions. Also, our 
band structure calculations on Pr$_{0.75}$Sr$_{0.25}$MnO$_3$ show the 
possible existence of half-metallicity in this system.

\end{abstract} 
\pacs{79.60.-i, 75.47.Gk, 71.30.+h} 
\keywords{Photoemission spectroscopy, CMR, Insulator-metal transitions}
\maketitle

\section{\bf INTRODUCTION}

The near Fermi level electronic structure of colossal magnetoresistance 
(CMR) materials \cite{tokurabook} (RE$_{1-x}$AE$_{x}$MnO$_3$, RE $=$ rare 
earth, AE $=$ alkaline earth) have been attracting much interest recently, 
particularly due to their importance in understanding the mechanism behind 
the insulator-metal (IM) transitions in these oxides. The ground state 
properties of these manganites are dominated mainly by the competition 
among four independent interactions: the magnetic interaction between the 
Mn $e_g$ electron spins, the electron-phonon coupling, the electronic 
repulsion and the kinetic energy of the carriers. The behavior of doped 
manganites for $x < 0.5$ and $x > 0.5$ is drastically different and 
depends strongly on the competition among the above four
interactions. These distorted perovskites undergo a transition from a
ferromagnetic metallic (FMM) state to an antiferromagnetic insulating 
(AFMI) state with the doping of charge carriers. However, for a 
commensurate fraction of the carrier concentration, the FMM state is 
easily suppressed by the formation of different types of charge-ordering
(CO). For example, CO in narrow one-electron bandwidth materials like
Pr$_{0.5}$Ca$_{0.5}$MnO$_3$ and La$_{0.5}$Ca$_{0.5}$MnO$_3$ is 
charge-exchange (CE) type checkerboardlike. On the other hand, CO in
wide one-electron bandwidth material like Pr$_{0.5}$Sr$_{0.5}$MnO$_3$ is 
``stripe-like"\cite{kajimoto}. The FMM to CO insulating transition is
caused by a spin and orbital-ordering of the carriers.

There have been several experimental investigations on the valence-band
electronic structure of manganites\cite{sarma1,park,dalai}. Recent 
photoemission studies have highlighted the importance of the subtle 
changes in the density of states (DOS) at the Fermi level (E$_F$) in 
understanding the nature of the IM transitions of these 
manganites\cite{chainani,saitoh1,sekiyama,kang,ebata1,ebata2}. Many of the 
theoretical models, particularly those based on electronic phase 
separation also emphasize on the importance of these 
states\cite{moreo1,moreo2,aliaga}. Chainani {\it et al.}\cite{chainani}
have shown that some of the DOS very close to E$_F$ get shifted to higher 
binding energies ($\sim$ 1.2 eV from E$_F$) and consequently a CO gap of
$\sim$ 100 meV opens up at the transition. These authors have remarked 
that such spectral weight transfer could be a generic feature of CO gap in 
manganites. Similar studies on the CO compositions of
Pr$_{1-x}$Ca$_{x}$MnO$_3$ are also reported\cite{ebata1,ebata2}. Although,
the mechanism of this CO gap is still not clear, the large energy scale
over which the spectral weight get transferred shows the possibility of
Coulombic origin. From the point of view of different theoretical models,
it is quite likely that the pseudogap (PG) associated with the IM
transitions, observed in double layered manganites by Saitoh {it et
al.}\cite{saitoh1} has the same origin as the CO gap. Such a generality 
in the shifts of near E$_F$ DOS is interesting to be probed. Apart from
the photoemission studies, PGs associated with the IM transitions of
manganites were confirmed also using scanning tunneling spectroscopic
studies\cite{mitra1,mitra2}. 

In order to understand the nature of the near E$_F$ spectral weight shifts
in systems with no CO, we have studied some of the compositions of the
Pr$_{1-x}$Sr$_x$MnO$_3$. The electrical and magnetic phase diagram of this
system has been published earlier\cite{martin1}. Different structural and
magnetization studies\cite{martin1,zirak,knizek,kawano} have shown that
there exists no CO in any of the compositions in the 
Pr$_{1-x}$Sr$_x$MnO$_3$ system. However, a neutron diffraction study
by Kajimoto {\it et al.}\cite{kajimoto} showed that the 
Pr$_{0.5}$Sr$_{0.5}$MnO$_3$ sample had an anisotropic 
``stripe-like" CO. Below the Curie transition (T$_C$), this system is FMM
for $0.25 < x < 0.5$, A-type AFMI for $0.5 < x < 0.7$ and C-type AFMI for
$0.7 < x < 0.9$. For identifying the nature of the bands contributing to
the near E$_F$ DOS, we studied one of the relevant compositions of the
Pr$_{1-x}$Sr$_x$MnO$_3$ system using the theoretical band structure
calculation method based on local spin-density approximation incorporating
the electron correlation (LSDA$+$U). The main part of this paper describes
the results of our high resolution photoelectron spectroscopic (PES) study
of the near E$_F$ states of some compositions of this system. Our studies
show a shift in the near E$_F$ spectral weight displaying a PG
behavior. As per our knowledge, there has been no study reported earlier
on the PG behavior of cubic manganites with no CO, using photoelectron
spectroscopy.

\section{\bf EXPERIMENT and CALCULATIONS}

Polycrystalline samples of the Pr$_{1-x}$Sr$_x$MnO$_3$ system were
prepared by solid state reactions\cite{martin1}. Purity and cationic
homogeneity of all the samples were systematically checked by electron
diffraction coupled with energy dispersive spectroscopy 
analysis. Resistivity and magnetic behavior of the samples have been
studied using the four probe technique and SQUID magnetometry. Details of
the sample preparation, characterization and structural studies are 
published elsewhere along with a phase
diagram\cite{martin1,martin2,hervieu}. 

Angle integrated ultraviolet photoemission measurements were performed
using an Omicron mu-metal UHV system equipped with a high intensity
vacuum-ultraviolet source (HIS $13$) and a hemispherical electron energy
analyzer (EA $125$ HR). At the He $I$ ($h$ $\nu$ = $21.2$ eV) line, the
photon flux was of the order of $10^{16}$ photons/sec/steradian with a
beam spot of $2.5$ mm diameter. Fermi energies for all measurements were
calibrated using a freshly evaporated Ag film on a sample holder. The
total energy resolution, estimated from the width of the Fermi edge, was
about $80$ meV for He $I$ excitation. All the photoemission measurements
were performed inside the analysis chamber under a base vacuum of 
$\sim$ $1.0$ $\times$ $10^{-10}$ mbar. The polycrystalline samples were
repeatedly scraped using a diamond file inside the preparation chamber
with a base vacuum of $\sim$ $1.0$ $\times$ $10^{-10}$ mbar and the
high-resolution spectra were taken within $1$ hour to avoid surface
degradation. The measurements were repeated to ensure reproducibility of
the spectra. For the temperature dependent measurements, the samples were
cooled by pumping liquid nitrogen through the sample manipulator fitted
with a cryostat. Sample temperatures were measured using a silicon diode 
sensor touching the bottom of the stainless steel sample plate.

\begin{figure}[t]
\vskip 1.0cm
\includegraphics[width=3.0in]{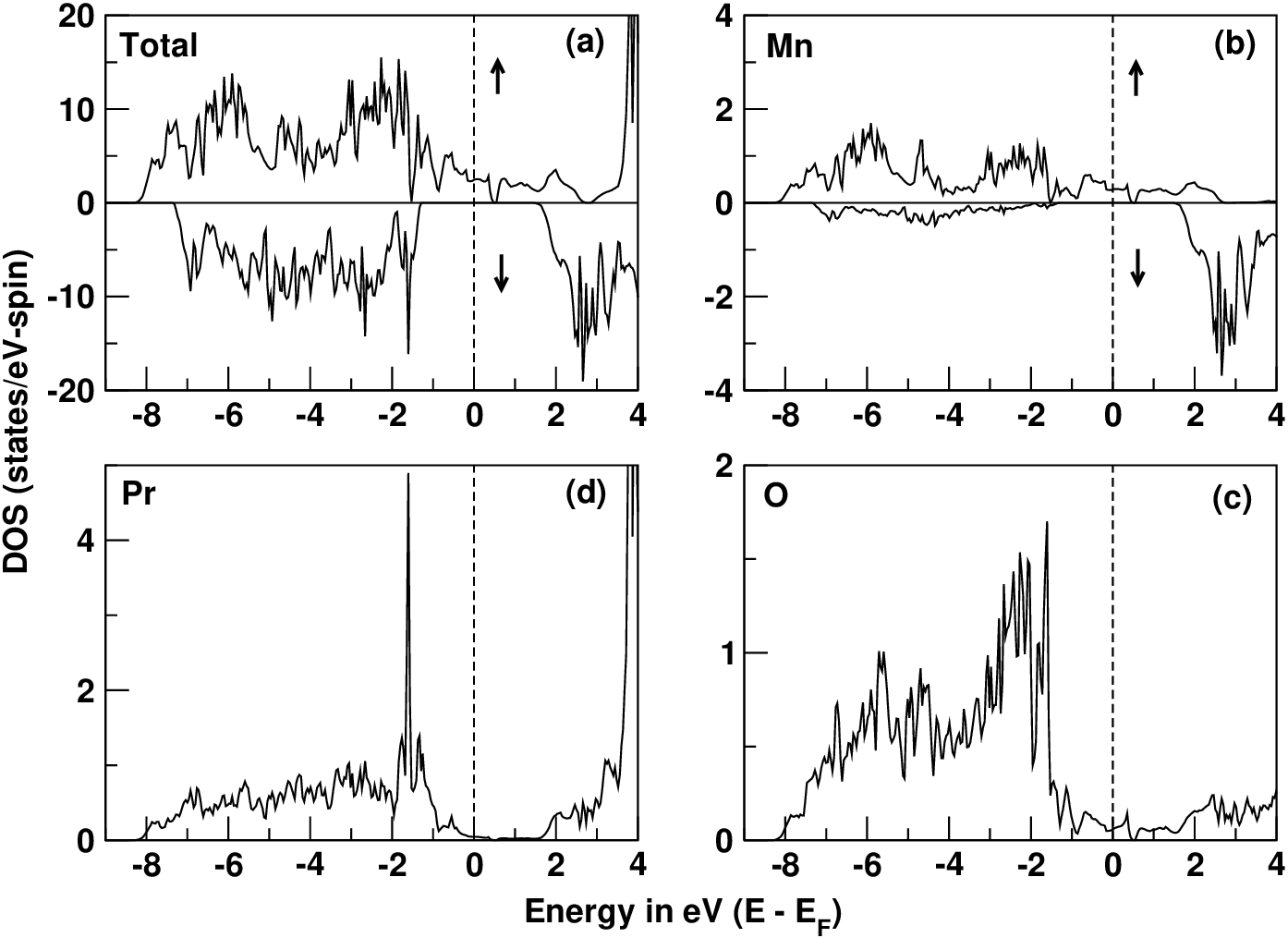}
\caption{\label{figure_1}Band structure of Pr$_{0.75}$Sr$_{0.25}$MnO$_3$ 
calculated using the LSDA + U method. (a) shows the total spin projected
DOS. (b), (c) and (d) show the site projected DOS. The total spin 
projected DOS shows the half-metallic behavior of this composition.}
\end{figure}

The LSDA$+$U band-structure calculations for 
Pr$_{0.75}$Sr$_{0.25}$MnO$_{3}$ were done for the ferromagnetic phase
within the frame work of tight-binding Linear Muffin-Tin Orbitals
(LMTO) method in the atomic sphere 
approximation\cite{ander1,ander2}. Though, this composition has a slightly
lower hole content ($x$) than the most important sample ($x$ = 0.35) of
the experimental part of this study, we have chosen it for the convenience
of calculation. Nevertheless, both these compositions show similar
electrical and magnetic properties as per the phase diagram of this
system\cite{martin1}. The crystal structure of 
Pr$_{0.75}$Sr$_{0.25}$MnO$_{3}$, we have used for the LSDA$+$U super-cell
calculation is that of the parent compound PrMnO$_3$ (space
group: Pbnm)\cite{zirak}. The unit cell taken was four formula unit. We
have substituted one of the Pr atoms with a Sr atom. The exchange term $J$
and the correlation term $U$ for Pr $4f$ state are $0.95$ eV and $7$ eV
respectively and that for Mn $3d$ is $0.87$ eV and $4$ eV 
respectively. The k-mesh used for the self-consistent calculation is
(10$\times$10$\times$10).

\section{\bf RESULTS AND DISCUSSION}

In Fig. 1 (a) we present the total DOS of Pr$_{0.75}$Sr$_{0.25}$MnO$_3$
calculated using the LSDA + U method. We have chosen this composition for
a comparison with our experimental results and the issues addressed in
this paper. The finite DOS near the E$_F$ in Fig. 1(a) shows that
Pr$_{0.75}$Sr$_{0.25}$MnO$_3$ is metallic, consistent with the phase
diagram reported earlier\cite{martin1}. The near E$_{F}$ states are
dominated by the Mn and O. From their individual DOS (Fig.1(b) and 
1(c)) it is quite clear that there is a strong hybridization between the
Mn $3d$ and the O $2p$ states through out the valence band. Though, the Pr
$4f$ states appear at around $2$ eV from the E$_F$, the electronic
properties of this system are dominated by the Mn $3d$ states close to
E$_F$. The states closest to the E$_F$ are due to the Mn 3d $e_g$
orbitals. As per our knowledge there has been no earlier reports published
on the band structure calculations of this composition. However, the
results of our calculation are in agreement with the LMTO and LSDA 
calculations reported earlier on similar systems\cite{pickett,satpathy}. 
Figs. 1(b), 1(c) and 1(d) show the site projected DOS for 
Pr$_{0.75}$Sr$_{0.25}$MnO$_{3}$. Based on the total spin projected DOS
(Fig. 1(a)), we predict that Pr$_{0.75}$Sr$_{0.25}$MnO$_{3}$ could exhibit
a half-metallic behavior, though there have been no experimental evidence
to support this. The majority band is conducting whereas the minority band
is clearly insulating; the hallmark of half-metallicity. The minority Mn
band has very little population below E$_F$, consequently Mn atoms in
Pr$_{0.75}$Sr$_{0.25}$MnO$_{3}$ has a substantially high magnetic
moment. The calculated magnetic moment of individual Mn atom in
Pr$_{0.75}$Sr$_{0.25}$MnO$_{3}$ is 3.88 $\mu_{B}$. The total 
magnetic moment of the super-cell is 15 $\mu_{B}$. The integral
magnetic moment is the signature of the resulting 
half-metallicity. Half-metallicity in these compositions are important for
the understanding of CMR phenomena as well as their technological
applications in spintronics. A detailed study of the half-metallic 
compositions of this system is being published elsewhere\cite{monodeep}.

\begin{figure}[t]
\vskip 1.0cm
\includegraphics[width=3.0in]{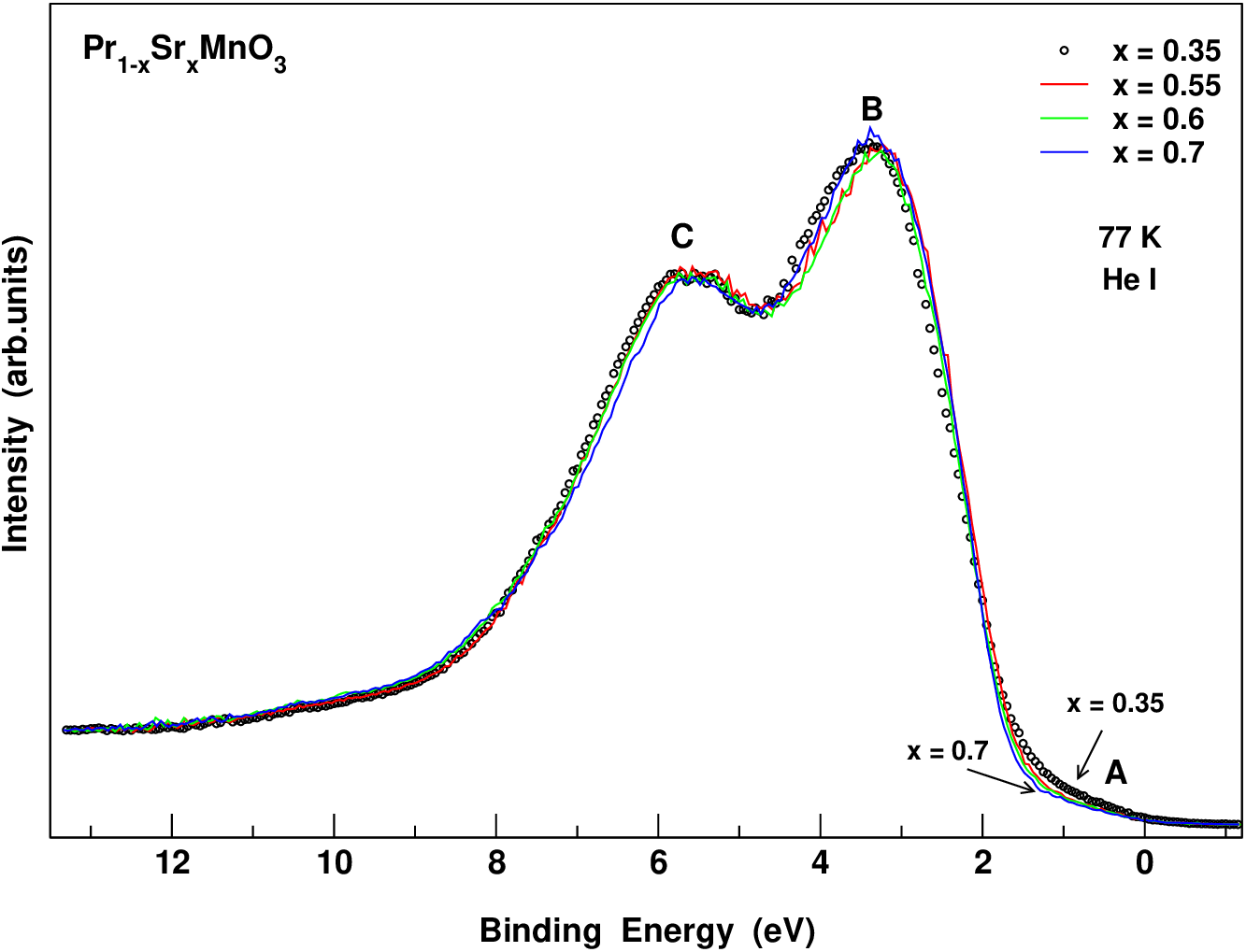}
\caption{\label{figure_2}(Color online) The experimental valence band
photoemission spectra of the Pr$_{1-x}$Sr$_x$MnO$_3$ samples with $x =
0.35$, 0.55, 0.6 and 0.7 taken at $77$ K (h$\nu = 21.2$ eV).}
\end{figure}

Figure 2 shows the experimental valence band photoemission spectra of the 
Pr$_{1-x}$Sr$_x$MnO$_3$ samples with $x$ = 0.35, 0.55, 0.6 and 0.7 taken
at $77$ K using UV photons of energy $21.2$ eV. The spectra exhibit three
features $A$ ($0.7$ eV), $B$ ($3.3$ eV) and $C$ ($5.5$ eV) derived mainly
from the hybridized Mn $3d$ - O $2p$ states in agreement with our
calculations shown in Fig. 1. The feature $A$ is due to Mn $e_g$ spin-up
states while the feature $B$ has a strongly mixed character of Mn $3d$
$t_{2g}$, O $2p$ and Pr $4f$ states. The feature $C$ has a dominant O $2p$
character with a small Mn $3d$ contribution. These assignments are 
consistent with the earlier reports on related manganites using
photoemission 
spectroscopy\cite{dalai,chainani,sekiyama,kang,ebata1,saitoh2}. Due 
to the large ionization cross-section, at this photon energy, 
the O $2p$ contribution to the spectra is high compared to the Pr $4f$ and
Mn $3d$. As one can see from the figure, the intensity of feature
($A$) decreases systematically with increasing Sr content ($x$). The
figure also shows that there are no drastic changes in the features $B$
and $C$ as a function of composition, which is again consistent with
earlier experiments\cite{sarma1,park}. It is to be noted that the
Pr$_{0.65}$Sr$_{0.35}$MnO$_3$ sample has a significantly high 
spectral intensity near the E$_F$, depicting the metallic nature of this 
composition.

\begin{figure}[t]
\vskip 1.0cm
\includegraphics[width=3.0in]{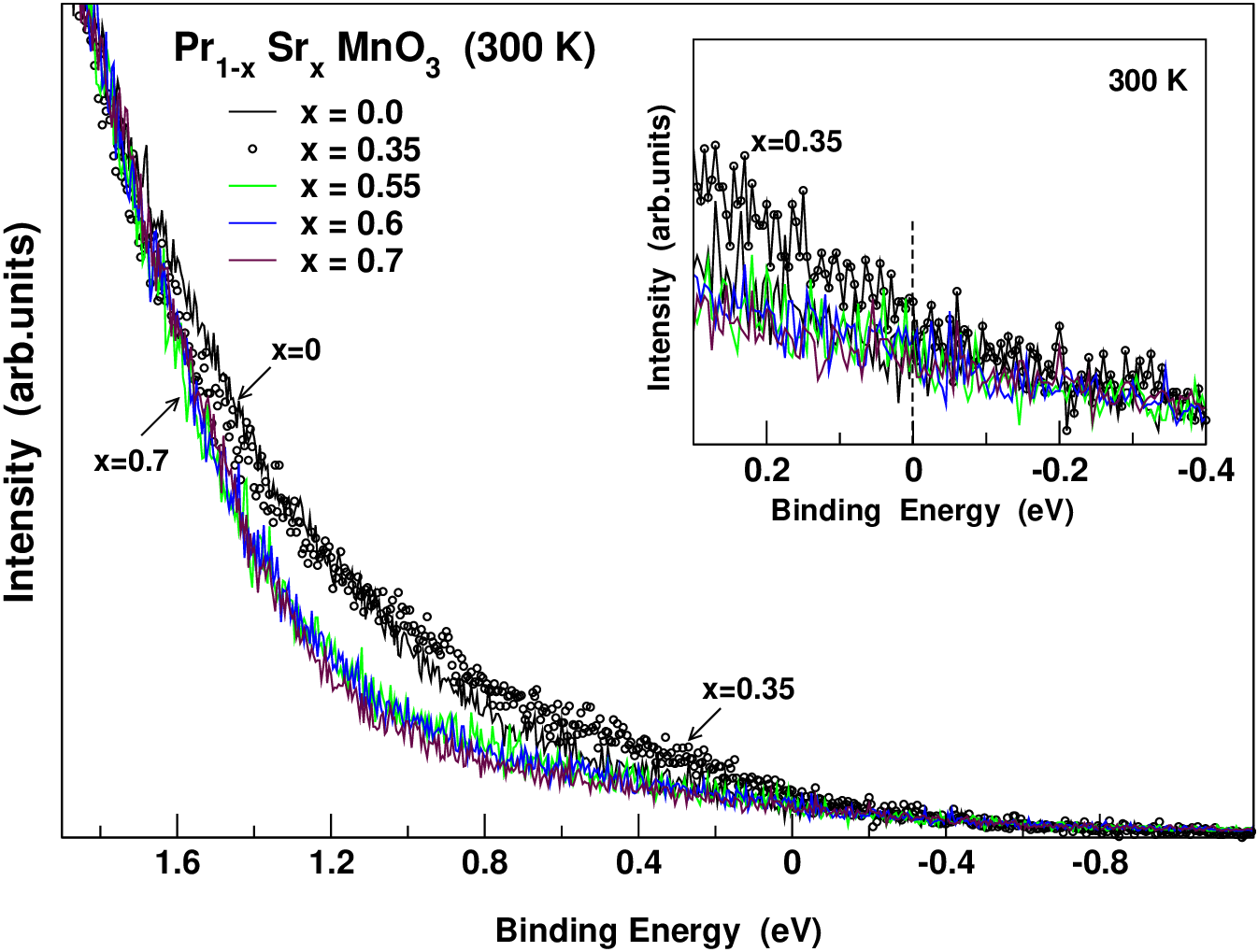}
\caption{\label{figure_3}(Color online) Doping dependent near E$_F$
high-resolution photoemission spectra of Pr$_{1-x}$Sr$_x$MnO$_3$ compound
taken at $300$ K. All the spectra have been normalized to the integrated
intensities in the energy region displayed here. The inset shows the
spectral changes in the vicinity of E$_F$ at $300$ K.}
\end{figure}

\begin{figure}[t]
\vskip 1.0cm
\includegraphics[width=3.0in]{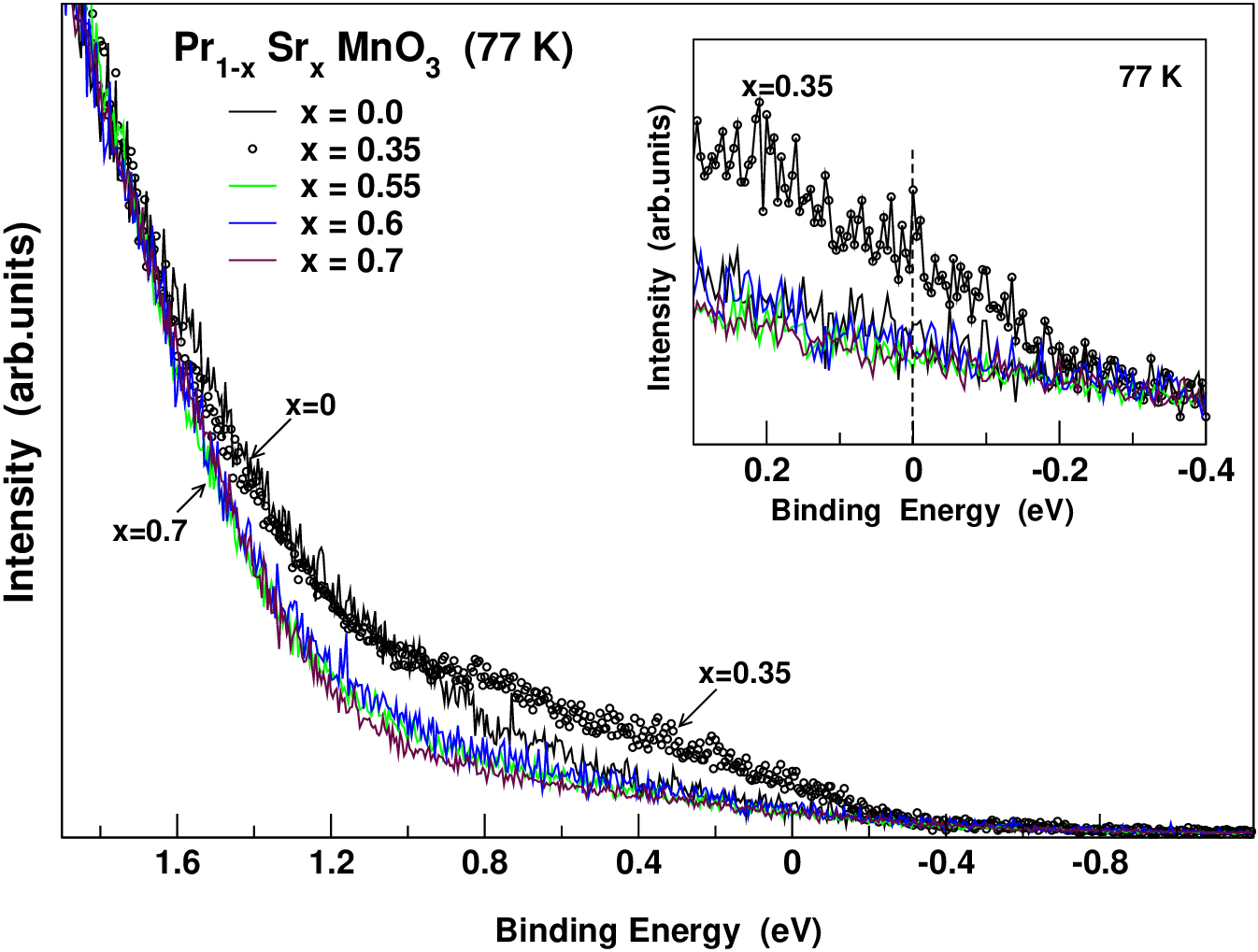}
\caption{\label{figure_4}(Color online) The near E$_F$ photoemission 
spectra of Pr$_{1-x}$Sr$_x$MnO$_3$ compound taken at $77$ K. The inset
shows the spectral changes in the vicinity of E$_F$ at $77$ K.}
\end{figure}

In this paper we concentrate on these near E$_F$ electronic states
responsible for most of the electrical and magnetic properties of
Pr$_{1-x}$Sr$_x$MnO$_3$ system. In order to study these states, we have
performed our temperature dependent high resolution PES measurements on
the Pr$_{1-x}$Sr$_x$MnO$_3$ compounds. Figure 3 shows the near E$_F$
spectra for various concentrations taken at $300$ K. The spectra are
normalized by the integrated intensity below E$_F$ and all above $1.7$
eV. One can see that the near E$_F$ spectral weight gradually decreases as
we go from the parent compound ($x = 0.0$) to $x = 0.7$, except in 
case of the Pr$_{0.65}$Sr$_{0.35}$MnO$_3$. Though, this systematic
decrease is expected as the average $e_g$ electron number decreases with
increasing $x$, the $x = 0.35$ sample shows a comparatively high spectral
weight near E$_F$, despite its insulating nature at this
temperature. Figure 4 presents the high resolution PES spectra of all the
samples taken at $77$ K. At this temperature the 
Pr$_{0.65}$Sr$_{0.35}$MnO$_3$ sample shows FMM properties and all other 
compositions are insulating\cite{martin1}. Again, these low temperature
spectra show almost no change in the near E$_F$ spectral weight compared
to that taken at room temperature, except for the $x = 0.35$ which shows a
distinct increase in these states. This increase indicates the metallic
behavior of Pr$_{0.65}$Sr$_{0.35}$MnO$_3$ at $77$ K. The insets of Fig. 3
($300$ K) and Fig. 4 ($77$ K) provide a closer view of the intensity
changes in a narrow energy region near E$_F$. A comparison of the two
insets will further demonstrate that except the $x = 0.35$ all other
samples show negligible change with lowering of temperature from $300$ K
to $77$ K.

\begin{figure}[t]
\vskip 1.0cm
\includegraphics[width=3.0in]{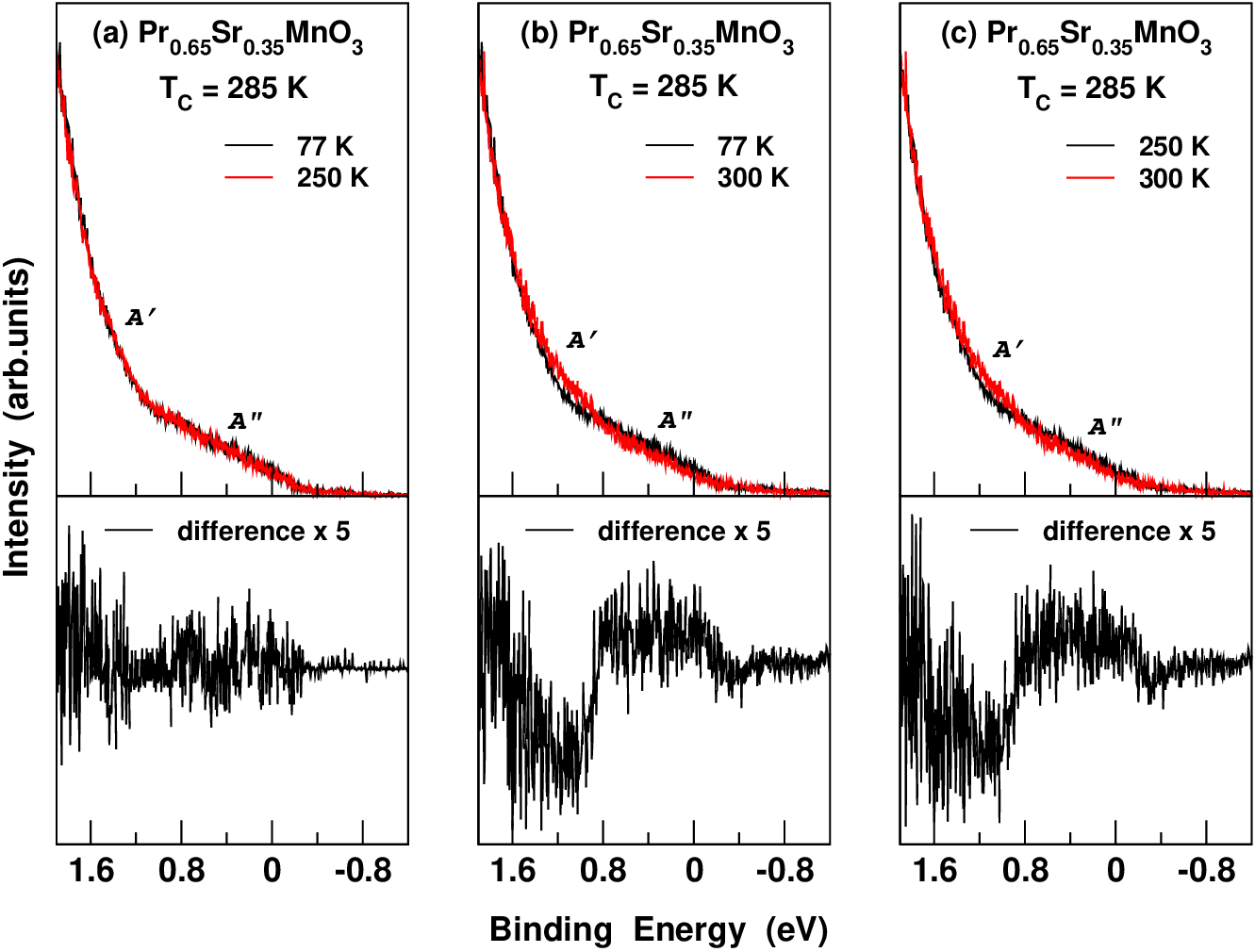}
\caption{\label{figure_5}(Color online) The near E$_F$ high-resolution
spectra of Pr$_{0.65}$Sr$_{0.35}$MnO$_3$ taken as a function of
temperature (a) at T$ = 77$ and $250$ K, indicating negligible changes in
the FMM phase with temperature, (b) at T$ = 77$ and $300$ K, and (c) at T$
= 250$ and $300$ K, across the FMM-PMI transition at T$_C = 285$ K
showing a spectral weight transfer with decreasing temperature. Lower
panels show the difference in spectra magnified by a factor of $5$.} 
\end{figure}

We have further analyzed this near E$_F$ spectral weight change in case of 
the Pr$_{0.65}$Sr$_{0.35}$MnO$_3$ sample in order to understand its 
temperature dependence and origin. In Fig. 5 we compare the near E$_F$
spectra obtained at $77$ K (well below the paramagnetic insulating 
(PMI) to FMM transition, T$_C$), $250$ K (below the T$_C$) and $300$ K
(above the T$_C$). The two spectra taken at $77$ K and $250$ K plotted
together (Fig. 5(a)) show little change in their near E$_F$ spectral
weight as a function of temperature, whereas the $250$ K and $300$ K
spectra (Fig. 5(c)) show a shift of spectral weight from the feature
marked A" to A' as we go across the FMM-PMI transition. Such a shift is
visible between the $77$ K and $300$ K spectra (Fig. 5(b)) also. 
Interestingly, the magnitude of the spectral weight shift between A" to A'
remains the same for the $250$ K or $77$ K, showing that below T$_C$ no
further shift of spectral weight occurs with temperature. This shift of
spectral weight from A" to A' while going across the FMM to PMI state
shows the opening up of a gap at T$_C$. Since, the PMI state also shows a
finite amount of states at the E$_F$, we prefer to call this a PG. From
the difference spectra shown in the lower panels it is clear that this PG
occurs due to the shift of states over an energy scale $\sim$ $1.0$ eV. In
other words, $\sim$ $1.0$ eV (difference between A' and A") is required
for the delocalization of the charge carriers to form the FMM state from
the localized insulating state in this system. Now, looking back at Fig. 3
will reveal that the energy position of A' is the same as that of the
spectral bump shown by the parent compound (PrMnO$_3$) at $\sim$ $1.2$ eV
below E$_F$, which corresponds to the localized $e_g$ states in this
insulator. When the charge carrier concentration, $x$ is increased to
$0.35$ these localized $e_g$ states melt to the position of A" due to the
changes in the overlap of Mn $3d$ - O $2p$ orbitals, as a consequence of
the structural changes (Mn-O-Mn bond angles). This delocalization becomes
stronger below T$_C$ resulting in the metallic behavior of $x = 0.35$
sample. 

As mentioned earlier, there have been a few studies using PES on the CO
gap in similar 
systems\cite{chainani,sekiyama,kang,ebata1,ebata2}. Although, none of our 
Pr$_{1-x}$Sr$_x$MnO$_3$ samples show any CO at any temperature, the energy
scale involved in the shift of states from A' to A" in our spectra is
comparable to the value ($1.2$ eV) observed in a CO 
composition\cite{chainani}. This indicates that the PG observed
here and the CO gap are similar in nature, suggesting a common origin for
both. As we have seen in the Fig. 5 the spectral weight shift in the
Pr$_{0.65}$Sr$_{0.35}$MnO$_3$ sample from A' to A" across the T$_C$ ($300$ 
K to $250$ K) remains the same even with cooling down to $77$ K. This
shows there is no further temperature dependent delocalization of charge
carriers except at the T$_C$. Again, this is another similarity between
the CO gap and the PG, that both are abrupt at T$_C$ with no
temperature dependencies above or below\cite{chainani}. These similarities
may further suggest that a delocalization of the charge carriers and
thereby a shift of the near E$_F$ DOS is generic to the IM
transitions in these manganites, regardless of the presence or absence of
any CO. 

The shift of electron states from near E$_F$ to the energies above it, is
an important result in understanding the CMR effect. Moreo {\it et
al.} have shown that such PG can well be accounted in models based on
electronic phase separation\cite{moreo2}. In these models, the PG
behavior was found to be robust in systems with mixed-phase
characteristics, where FMM clusters exist in an insulating background. The
abrupt nature of the PG at T$_C$ found in our study also can be
understood, since these mixed phase tendencies are strong at temperatures
around T$_C$. Signatures of mixed phases are visible even at temperatures
above or below the T$_C$, in the form of the absence of a sharp Fermi 
edge even in FMM phases and the existence of a finite number of states at
the E$_F$ in the PMI state. Monte Carlo calculations\cite{moreo1} predict
that the PG at the chemical potential should be generic of mixed phases.

\section{conclusion}

Our studies using UV photoelectron spectroscopy show that the shifts of
DOS near the E$_F$ observed across the IM transitions in
compositions of the Pr$_{1-x}$Sr$_x$MnO$_3$ system with no CO, 
are abrupt at the T$_C$ and occur over an energy scale of
$\sim$ $1$ eV. The energy scale involved in this PG formation and
the abrupt nature of it near the T$_C$, strongly suggest that the CO gap
observed earlier in other manganites and the PG may indeed have
same origin. Further, we infer from our results as well as the results 
shown by Chainani {\it et al.}\cite{chainani} and Saitoh {\it et 
al.}\cite{saitoh1} that such shifts of the near E$_F$ DOS could be generic
to the IM transitions in the CMR compounds, regardless of the presence or
absence of any charge ordering. The PG observed associated with the IM
transition in Pr$_{0.65}$Sr$_{0.35}$MnO$_3$, shows the delocalization of
the charge carriers involved. Our results could be understood within the
framework of models based on the electronic phase separation where it has
been shown that the PG is a common behavior of CMR compositions with 
mixed-phase character. Our theoretical band structure calculations using
LSDA + U method on Pr$_{0.75}$Sr$_{0.25}$MnO$_3$, which is close in
character to the $x = 0.35$ sample, show the possible existence of
half-metallicity in this system.

\end{document}